# Redprint: Integrating API Specific "Instant Example" and "Instant Documentation" Display Interface in IDEs

*Anant Bhardwaj*
Stanford University
Gates Computer Science
Stanford, CA 94305
anantb@cs.stanford.edu

*Dave Luciano*
Stanford University
Gates Computer Science
Stanford, CA 94305
dluciano@stanford.edu

*Scott R. Klemmer*
Stanford University
Gates Computer Science
Stanford, CA 94305
srk@cs.stanford.edu

## ABSTRACT

Software libraries for most of the modern programming languages are numerous, large and complex. Remembering the syntax and usage of APIs is a difficult task for not just novices but also expert programmers. IDEs (Integrated Development Environment) provide capabilities like auto-complete and intellisense to assist programmers; however, programmers still need to visit search engines like Google to find API (Application Program Interface) documentation and samples. This paper evaluates Redprint - a browser based development environment for PHP that integrates API specific "Instant Example" and "Instant Documentation" display interfaces. A comparative laboratory study shows that integrating API specific "Instant Example" and "Instant Documentation" display interfaces into a development environment significantly reduces the cost of searching and thus significantly reduces the time to develop software.

**ACM Classification:** H5.2 [Information interfaces and presentation]: User Interfaces. - Graphical user interfaces.

## INTRODUCTION

Most of the modern programming languages have large software libraries. Programmers face a difficult task of remembering the name, syntax and usage of APIs.

Although most of the IDEs provide capabilities like auto-complete which helps programmers in quickly recalling the API name by typing a few initial characters and intellisense which helps programmers in recalling and validating the syntax of the API they are writing, these capabilities are of limited use to programmers because programmers still not remember the other details like the possible return values for an API – for example, the PHP API *int strcmp($str1, $str2)* may return 0, 1, -1 depending upon whether the function argument $str1 is equal to or greater than or less than the function argument $str2. Some functions modify the function arguments and return success or failure, while some return a new computed value. Some functions have dependencies – for example *fread()* requires a file handle returned from *fopen()*. These are the complexities which force programmers to look for API documentation and samples while writing or reading code.

For finding API documentation and sample code, programmers either visit the language reference website or they make API specific queries to search engines like Google. In our pilot study we found that for a typical programming session, API specific queries contribute for 43% (SD: 7.67) of the total search queries.

Programmers often look to borrow existing code from the web to reuse in their projects - there is a significant amount of time that is spent in understanding the borrowed code. Also, most of the software companies already have their own existing codebase which programmers are required to understand for writing new functionalities. In our pilot study, we found that while reading/understanding borrowed/existing code, the API specific queries contribute for 78% (SD: 11.98) of the total queries. While writing code from scratch, 23% (SD: 4.39) of the total queries were API specific.

The above observations form the basis of our research. We propose integrating two interfaces into IDEs, namely, "Instant Example" display interface and "Instant Documentation" display interface, to obviate the need of API specific queries and significantly reduce the cost of searching.

In Redprint, both these interfaces are always active and running in the background. As a user types anything in the editor window, (while writing code) the Redprint intellisense figures out the user's intended keyword/API and shows relevant examples and documentation in the Instant Example display window and the Instant Documentation display window respectively.

When a user is reading/understanding code, the Redprint intellisense looks for APIs and keywords at the line on which the cursor is placed and shows relevant examples and documentation in their respective windows (Fig. 1).

## RELATED WORK

Prior work has explored leveraging code examples from the web, inside integrated development environments. Blueprint is a search interface for Adobe Flex Builder which places search directly inside the development environment. However, it is task-specific i.e. it is oriented specifically towards finding example code. On the other hand, Redprint augments Blueprint to provide API specific examples and documentation on the fly to significantly reduce API specific searches.

## DESIGN

Task specific interfaces help programmers find example code. However, programmers still need to understand the APIs used in the borrowed code. Redprint separates example codes extracted from the web into two categories – API Specific and Task Specific. The API specific search interface monitors the user's cursor position, searches for relevant API specific examples/documentation in the background and displays them on the fly. To support task specific search similar to Blueprint, Redprint also provides a search box and a hot key (Ctrl + Space).

## EXPERIMENT

We conducted a controlled lab study with 25 participants who had prior experience with PHP.

We created two setups

1. Customized Eclipse (CE) setup with only a task specific search interface (similar to Blueprint).
2. Redprint with task specific and API specific search interfaces.

We randomly assigned one of the two setups to each participant.

The laboratory study evaluated two hypotheses:

**H1:** Programmers using Redprint setup will make significantly lesser number of search queries.

**H2:** Programmers using Redprint setup will complete the task faster.

The lab study comprised two tasks. The first task was to understand a given PHP program and the second task was to write a PHP program. All the participants were given the same tasks.

To conclude the study we asked the participants a few questions about their experience with Redprint "Instant Example" and "Instant Documentation" display interfaces.

## RESULTS

### Number of Search Queries:

Participants with the Redprint setup made significantly lesser number of search queries in both the tasks. For the reading/understanding task, average number of search queries for the Redprint setup was 2.67 vs. 8.64 for the CE setup (p-value = $7.1 \times 10^{-18}$). For the writing task, average number of search queries for the Redprint setup was 3.1 vs. 5.7 for the CE setup (p-value = $5.7 \times 10^{-4}$).

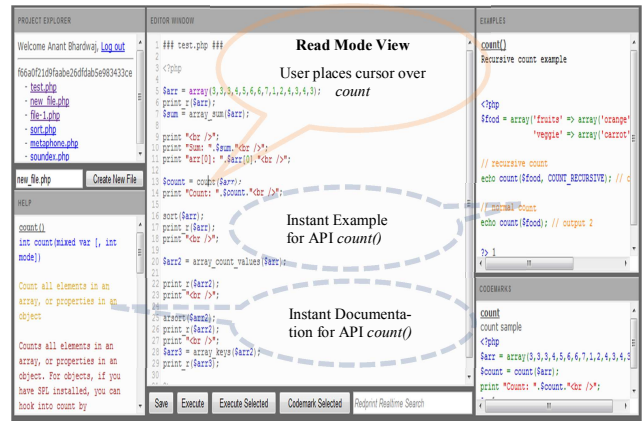

Figure 1: A typical reading/understanding mode view of Redprint - User's cursor is placed on line 13. Redprint intellisense recognizes count() API and displays examples and documentation for count() in "Instant Example" and "Instant Documentation" display interface respectively

### Time to complete the tasks

Participants with the Redprint setup took lesser time on an average. For the reading task, participants took on an average, 6.7 minutes in the Redprint setup vs. 11 minutes in the CE setup (p-value = 0.058). For the writing task, participants took on an average, 10.2 minutes in the Redprint setup vs. 13.1 minutes in the CE setup (p-value = 0.117).

### Post experiment Survey:

100% of the participants who worked on the Redprint setup responded that "instant API specific examples/documentation" helped them complete their tasks faster, out of which 81% reported a "very significant difference".

### CONCLUSION

To reduce the cost of searching, we proposed "Instant Example" and "Instant Documentation" interfaces. We then evaluated the Redprint IDE which integrates the abovementioned interfaces. In evaluating Redprint, we found that participants with instant enabled setup made significantly lesser number of search queries and were on an average, faster in completing their tasks.